\documentclass{iau}
\usepackage{graphicx}
\usepackage{color}  
\usepackage{hyperref}
\usepackage{graphicx,epstopdf}

\newcommand{\arcsec}{\mbox{$^{\prime\prime}$}}
\newcommand{\aap}{\textit{A\&A}}
\newcommand{\apj}{\textit{ApJ}}
\newcommand{\apjl}{\textit{ApJ Letters}}
\newcommand{\apjs}{\textit{ApJ Supplement Series}}

\newcommand{\araa}{\textit{ARAA}}
\newcommand{\zap}{\textit{Zeitschrift fuer Astrophysik}}

\title[The dense galactic environments of the Milky Way] 
{The dense galactic environments of the Milky Way}

\author[Q. Nguyen-Luong et al.]
{Quang Nguyen-Luong$^{1,2,3}$, Neal Evans$^{4,2}$, Kee-Tae Kim$^{2}$, Hyunwoo Kang$^{2}$ and DEGAMA survey} 

\affiliation{
$^1$IBM Canada, 120 Bloor Street East, Toronto, ON, M4Y 1B7, Canada, \\
$^2$Korea Astronomy and Space Science Institute, Yuseoung, Daejeon 34055, Korea, \\
$^3$Visiting researcher at the Graduate School of Natural Sciences, Nagoya City University, Japan\\
$^4$Department of Astronomy, The University of Texas at Austin, 2515 Speedway, Stop C1400, Austin, TX 78712-1205, USA
\\email: \tt{quang.nguyen@ibm.com}}

\pubyear{2019}
\volume{345}  
\jname{Origins: from the Protosun to the First Steps of Life}
\editors{Bruce G. Elmegreen, L. Viktor T\'oth, Manuel G\"udel, eds.}
\begin{document}

\maketitle

\begin{abstract}
Star formation takes place in the dense gas phase, and therefore a simple dense gas and star formation rate relation has been proposed. With the advent of multi-beam receivers, new observations show that the deviation from linear relations is possible. In addition, different dense gas tracers might also change significantly the measurement of dense gas mass and subsequently the relation between star formation rate and dense gas mass. We report the preliminary results the DEnse GAs in MAssive star-forming regions in the Milky Way (DEGAMA) survey that observed the dense gas toward a suit of well-characterized massive star forming regions in the Milky Way. Using the resulting maps of HCO$^{+}$ 1--0, HCN 1--0, CS 2--1, we discuss the current understanding of the dense gas phase where star formation takes place.
\keywords{stars: formation, ISM: clouds, ISM: structure, (ISM:) evolution, Galaxy: evolution}

\end{abstract}

\firstsection 
\section{Introduction}
We perform a survey of DEnse GAs in MAssive star-forming regions in the Milky Way (DEGAMA) survey to study the distribution of dense gas in molecular clouds and its role in forming stars. DEGAMA focusses on a larger sample that is sensitive to all gas above a column density threshold of 10$^{22}$\,cm$^{-2}$ with the goal of improving the understanding of how dense gas is formed and the relation between dense gas and star formation. The threshold column density of $N_{\mathrm{H}_2}\sim 10^{22}$\,cm$^{-2}$, which might correspond to a volume 
density of $1\times 10^{4}$\,cm$^{-3}$, is chosen 
because this gas directly builds massive star forming regions and forms lower mass stars (i.e., \cite[Onishi et al. 1998]{onishi98}, \cite[Andre et al. 2010]{andre10}).

At the extragalactic scale, we seek to understand the relationship between stellar 
density and gas density that was put forward by 
\cite[Thackeray (1948)]{thackeray48} and \cite[Van den Bergh (1957)]{vandenbergh57}), and crystallized in a relationship between the observable surface density 
of SFR, $\Sigma_{SFR}$, and the surface density of gas, $\Sigma{_{gas}}$, 
as: $\Sigma_{SFR} = A~\Sigma{_{gas}^N}$ by \cite[Schmidt (1959)]{schmidt59} and \cite[Kennicutt (1998)]{kennicutt98}.
However, this relation is not scale-invariant, the power-law indexes depend on the 
size scales of the objects, as pointed out using the diffuse cloud tracer CO 1--0 and radio continuum luminosity (\cite[Nguyen Luong et al. 2016]{nguyen-luong16}).
These relationships may behave differently if one considers 
only dense gas tracers, as was suggested for extragalactic 
environment (\cite[Wu et al. 2010]{wu10}, \cite[Liu et al. 2016]{liu16}).
The linear relationship manifests in the 
extragalactic environments (\cite[Gao \& Solomon 2004]{gao04}) and in the Milky Way environments (\cite[Lada et al. 2012]{lada12}), but significant gaps remain between 
extragalactic and galactic star formation laws (\cite[Heiderman et al. 2010]{heiderman10}). 
Nevertheless, the linear relationship between dense gas and SFR shows 
real scatter in excess of observational uncertainties (\cite[Usero et al. 2015]{usero15}), 
so additional factors are at work beyond a simple ``more dense gas gives 
more star formation'' model.  

\begin{figure}[!htbp]
\begin{center}
 \includegraphics[width=15cm]{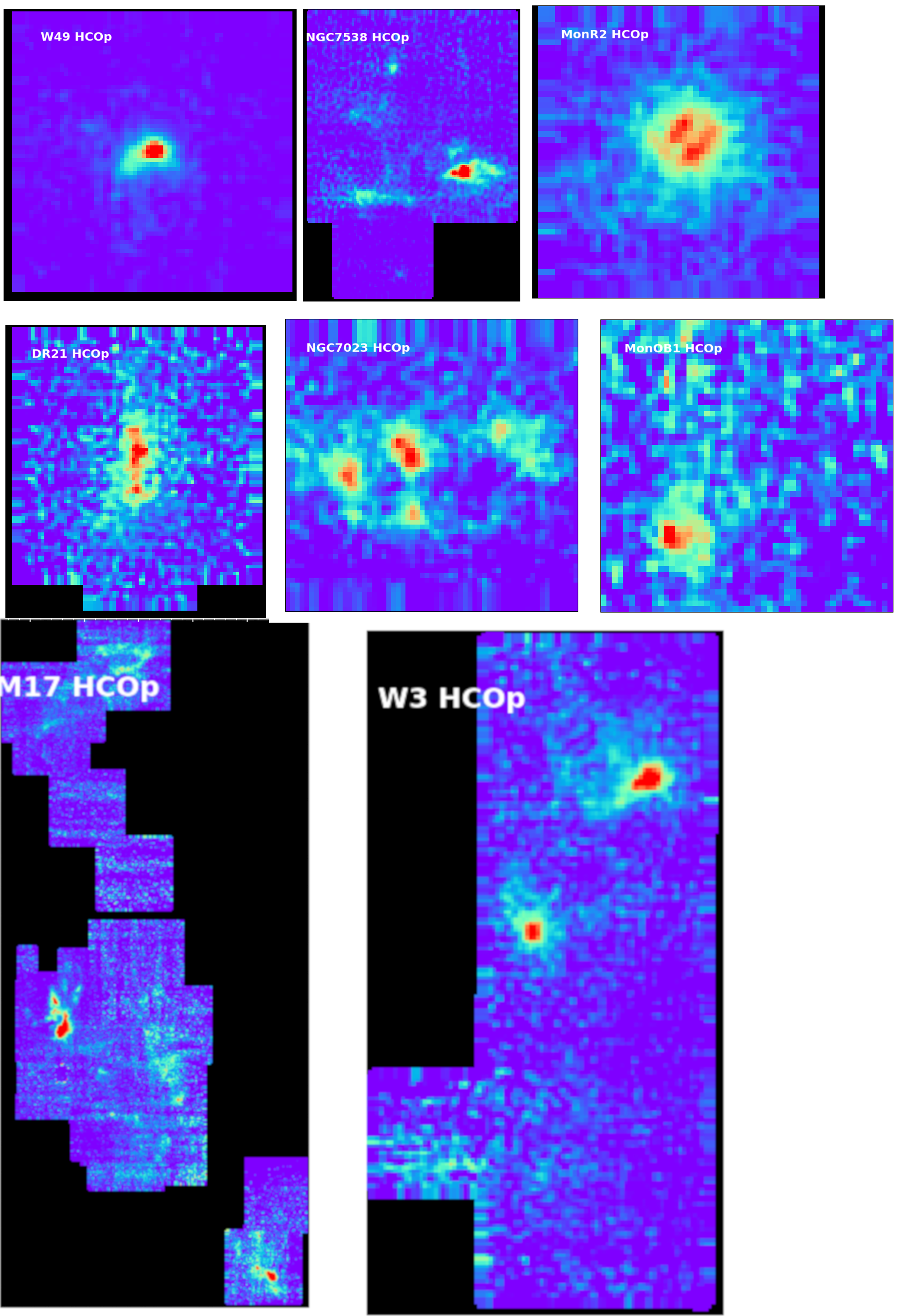}
 \caption{An overview of HCO$^{+}$ 1--0 integrated maps of DEGAMA targets.}
   \label{fig:int}
\end{center}
\end{figure}

At the extragalactic scale, we seek to understand the relationship between stellar 
density and gas density that was put forward by 
\cite[Thackeray (1948)]{thackeray48} and \cite[Van den Bergh (1957)]{vandenbergh57}), and crystallized in a relationship between the observable surface density 
of SFR, $\Sigma_{SFR}$, and the surface density of gas, $\Sigma{_{gas}}$, 
as: $\Sigma_{SFR} = A~\Sigma{_{gas}^N}$ by \cite[Schmidt (1959)]{schmidt59} and \cite[Kennicutt (1998)]{kennicutt98}.
However, this relation is not scale-invariant, the power-law indexes depend on the 
size scales of the objects, as pointed out using the diffuse cloud tracer CO 1--0 and radio continuum luminosity (\cite[Nguyen Luong et al. 2016]{nguyen-luong16}).
These relationships may behave differently if one considers 
only dense gas tracers, as was suggested for extragalactic 
environment (\cite[Wu et al. 2010]{wu10}, \cite[Liu et al. 2016]{liu16}).
The linear relationship manifests in the 
extragalactic environments (\cite[Gao \& Solomon 2004]{gao04}) and in the Milky Way environments (\cite[Lada et al. 2012]{lada12}), but significant gaps remain between 
extragalactic and galactic star formation laws (\cite[Heiderman et al. 2010]{heiderman10}). 
Nevertheless, the linear relationship between dense gas and SFR shows 
real scatter in excess of observational uncertainties (\cite[Usero et al. 2015]{usero15}), 
so additional factors are at work beyond a simple ``more dense gas gives 
more star formation'' model.  

With a critical density of $\sim10^{4}$\,cm$^{-3}$, HCO$^{+}$ 
1--0 (89.188 MHz) and HCN 1--0 (88.631 GHz), and CS 2--1 (97.980 GHz) is 
suitable to trace the total dense gas component in star forming regions. We observed these lines with the Taeduk Radio Astronomy Observatory (TRAO)  telescope \footnote{\url{https://radio.kasi.re.kr/trao/main_trao.php}} in the DEGAMA survey and report its first results in this paper. 

\section{Observations}
\label{sect:observation}
%
%
%
%
%
%
%
%

TRAO was established in October 1986 with the 13.7 meter Radio Telescope and recently equipped with the 16 pixels 4x4 SEQUOIA receiver. The 2nd IF modules with the narrow band and the 8 channels with 4 FFT spectrometers allow to observe 2 frequencies simultaneously within the 85--100 or 100--115 GHz bands for all 16 pixels of the receiver.
We carried out the DEGAMA mapping observations between December 2016 and December 2017. Observations were done in OTF mode and the native velocity resolution is less than 0.1 km/sec (15 kHz) per channel, and their full spectra bandwidth is 60 MHz. 
The telescope beam size is $\sim 50\arcsec$ at 100 GHz, and the main-beam efficiency $\eta_{\rm MB}$ is $\sim50\%$ at 100 GHz. 

Our target is a sample of massive star-forming cloud complexes in the 
Galaxy that present the entire massive star forming sequences from 
quiescent to active and evolved regions. This sample includes the 
massive star forming regions that are rather nearby (d$<5$\,kpc), so 
known star formation rates measurements from direct YSOs counting with 
Spitzer/WISE or Herschel. 
The sample includes the following sources that can be categorized in 
evolutionary stages, from quiescient to more active states:  M17, M16, DR21, IRAS05358, W3, MonOB1, NGC7538, NGC2264, W40, NGC7023 , MonR2, and Mon-OB1.

\section{Results}
\label{sect:results}
From a global perspective, dense gas tracers such as $L_{\rm HCO^+}$ or 
$L_{\rm HCN}$ and SFR tracers such as $L_{\rm IR}$ or $L_{\rm 
H_{\alpha}}$ are believed to correlate well with each other in 
log-log space (\cite[Gao \& Solomon .2004]{gao04}, \cite[Bussmann et al. 2008]{bussmann08}, \cite[Wu et al. 2010]{wu10}). The relations, 
however, are not identical for different transitions nor for different 
dense gas tracers (\cite[Krumholz \& Tan 2007]{krumholz07}, \cite[Narayanan 2008]{narayanan08}, \cite[Juneau et al. 2009]{juneau09}).  Moreover, 
intrinsic variation in this relationship suggests that simple correlations are inadequate for capturing the full range of star 
formation behavior (\cite[Usero et al. 2015]{usero15}).

We calculate the integrated line luminosity using equation developed in \cite[Solomon \& Vanden Bout (2005)]{solomon05} where
\begin{equation}
\frac{L'_{\rm line}}{{\rm K~km/s~pc^{2}}}
  = 23.5\times 10^{-6} \times \frac{D}{\rm kpc}^{2} \times \frac{A}{\rm 
\arcsec^{2}}\times \frac{I_{\rm line}}{\rm K~km/s}\, .
\end{equation}

\begin{figure}[!htbp]
\centering
$\begin{array}{ccc}
\includegraphics[angle=0,height=7cm]{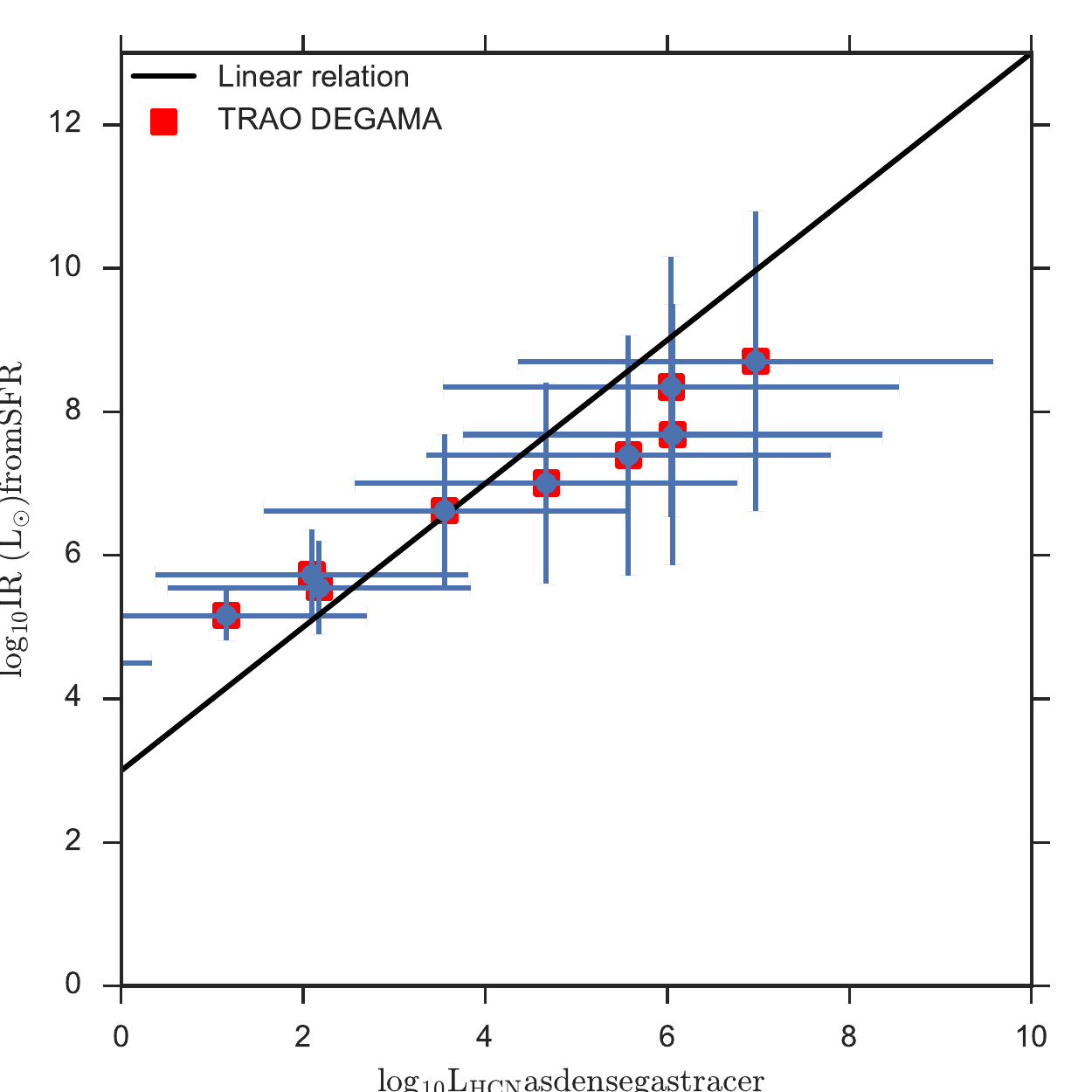} &
\includegraphics[angle=0,height=7cm]{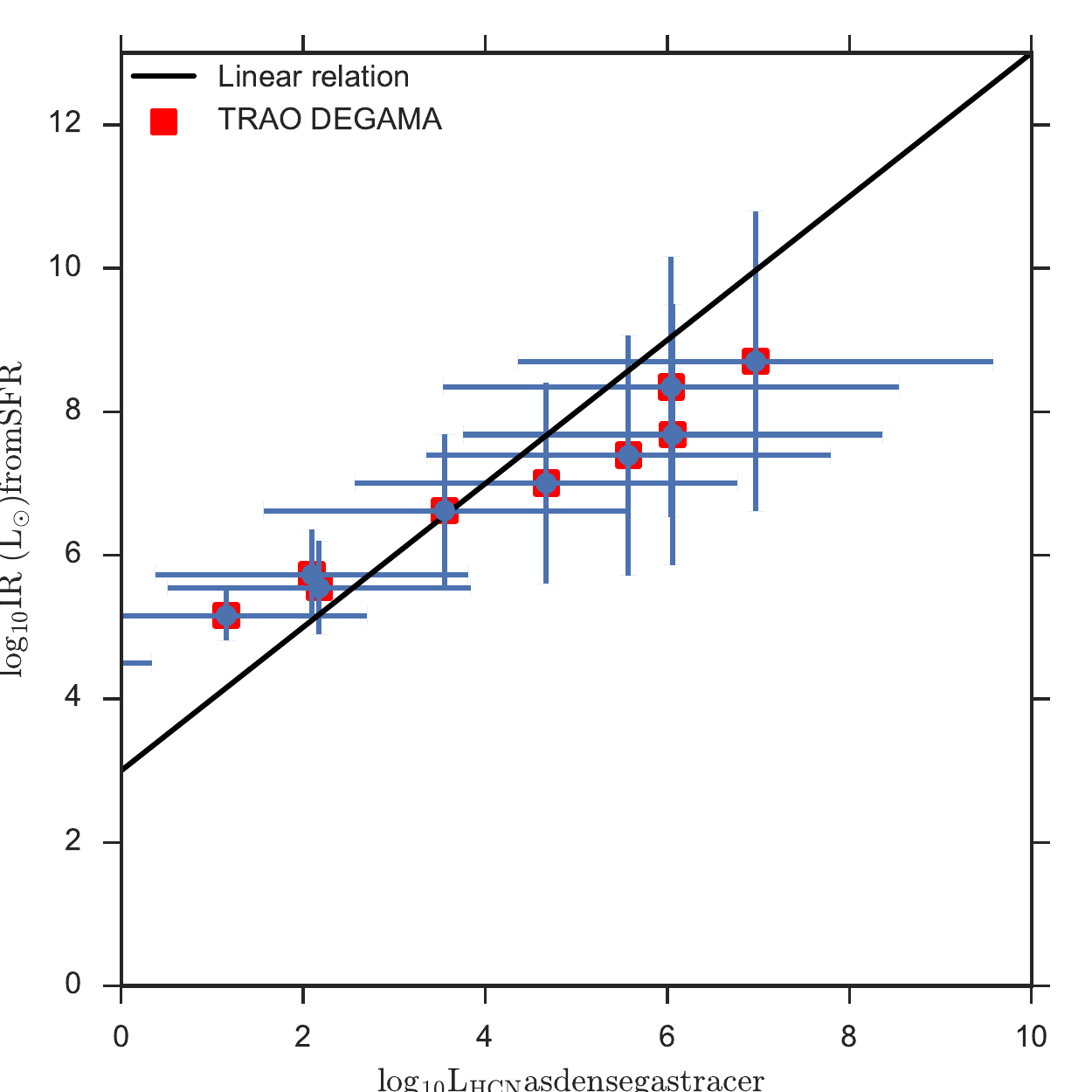}
\end{array}$
\caption{{\bf a:} 
Ratio $L_{\rm HCO^{+}}/L_{\rm HCN}$ as function of $L_{\rm HCN}$ 
luminosity, the horizontal line indicates where the average ratio 
$L_{\rm HCO^{+}}/L_{\rm HCN}=2$.  {\bf b}  $L_{\rm TIR}$ as 
function of $L_{\rm HCN}$ and the straightline is the linear function 
$\rm log(L_{\rm TIR}) = log(L_{\rm HCN})+3$ (\cite[Gao \& Solomon 2004]{gao04}, \cite[Wu et al. 2010]{wu10}).}
\label{fig:kslawcomparison}
\end{figure}

In DEGAMA survey, integrated line luminosity was derived from a collection of integrated maps of HCO$^{+}$ 1--0, HCN 1--0 and CS 2--1. Figure\,\ref{fig:int} shown an example of only HCO$^{+}$ 1--0 maps.  The ratio of $L_{\rm HCO^{+}}/L_{\rm HCN}$ from DEGAMA survey show that HCO$^{+}$ varies from cloud-to-cloud and vary around the average 
extragalactic ratio of 2 (Figure~\ref{fig:kslawcomparison}a). It might show that although having similar critical density, HCO$^{+}$  and HCN might trace quite different 
types of dense gas 
that 
When plotting on the 
$L_{\rm TIR}$-$L_{\rm HCN}$ plane, there is evidence that Galactic clouds 
do not follow a simple linear relation as the extragalactic clouds or 
the scatter is very large (Figure~\ref{fig:kslawcomparison}b).

\begin{figure}[!htbp]
\centering
$\begin{array}{ccc}
\includegraphics[angle=0,width=6cm]{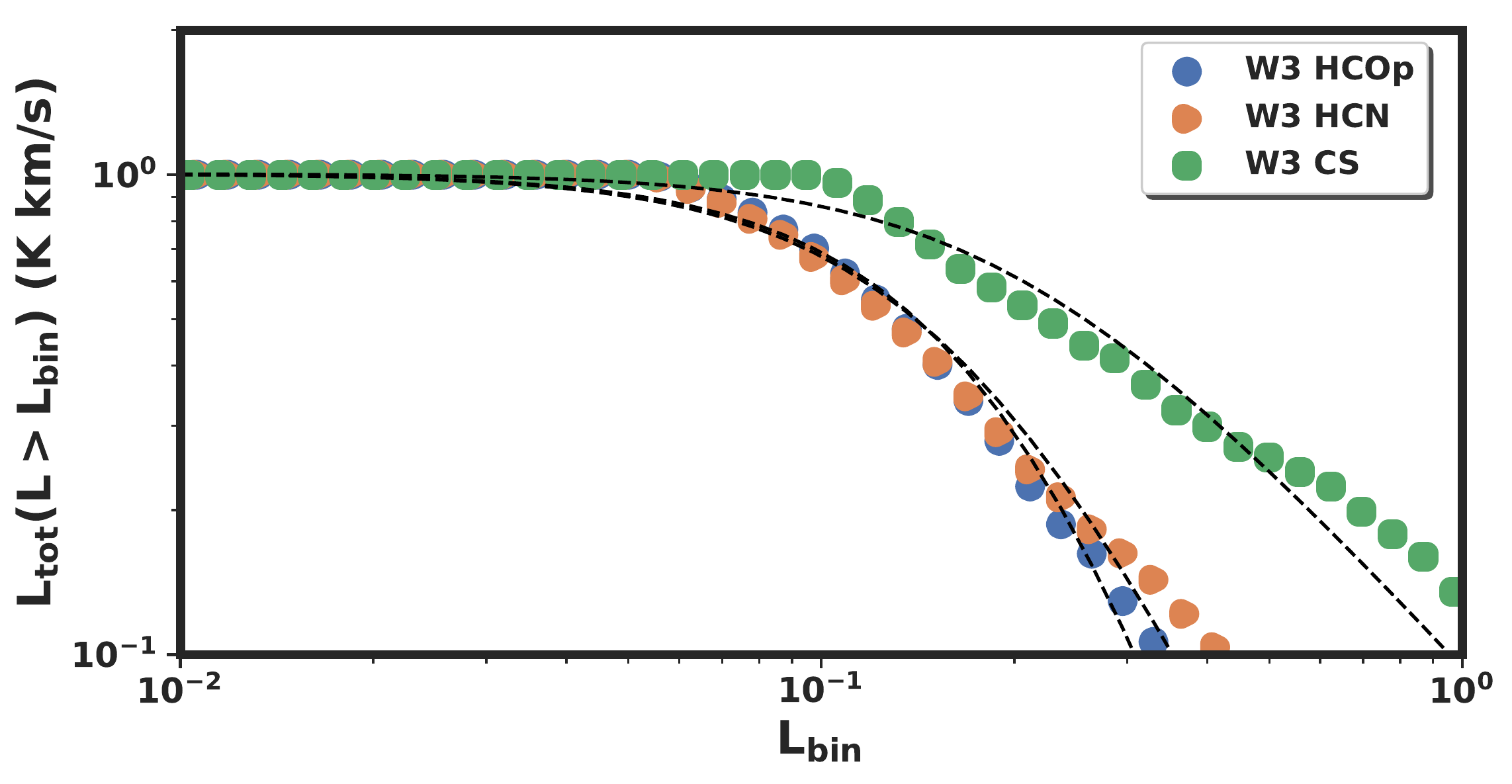} &
\includegraphics[angle=0,width=8cm,height=3.7cm]{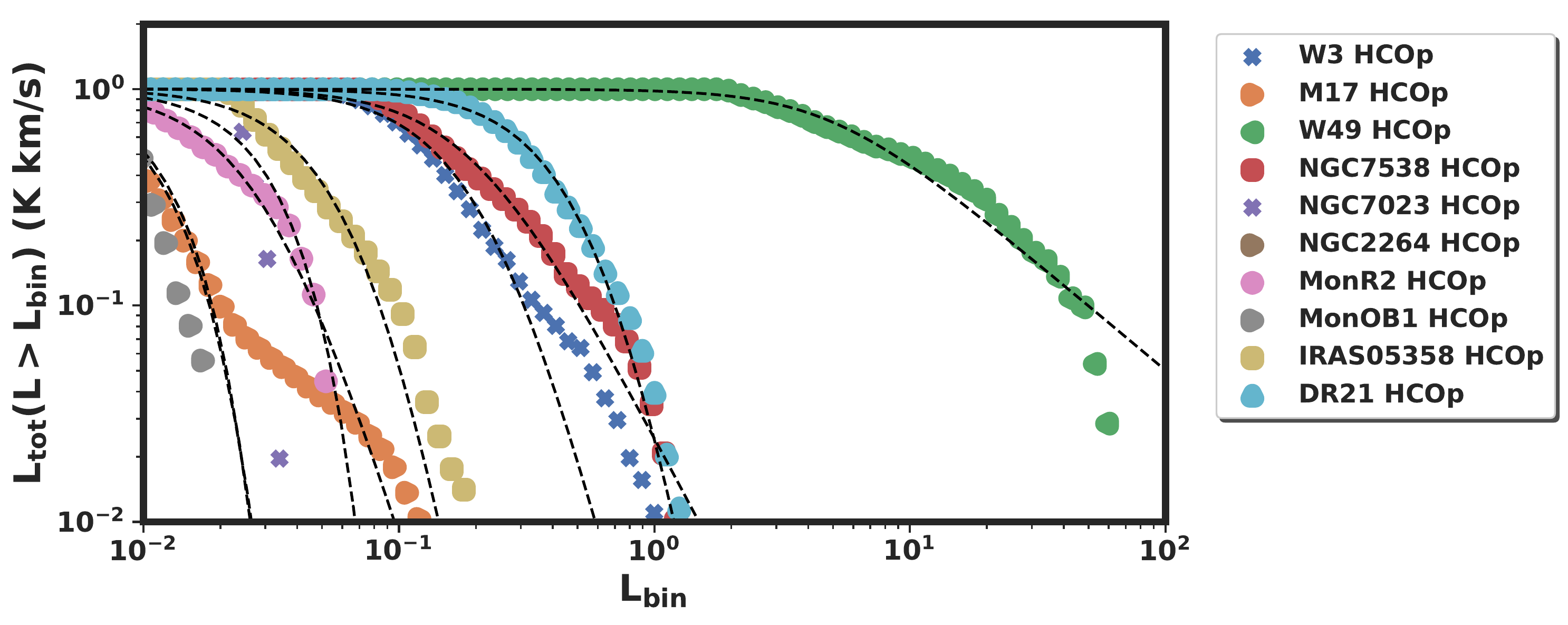}
\end{array}$
\caption{Normalized cumulative integrated line luminosity distributions of different tracers {\bf (a)} and normalized cumulative integrated line luminosity distributions of HCO$^{+}$ for different DEGAMA targets {\bf (b)}. }
\label{fig:ll2}
\end{figure}

Then, we divide the integrated line luminosities into a 100 bins and calculate the cumulative distributions (CDs) of the integrated line luminosities for each bin.  The results are plotted as functions of  luminosity bin in normalized forms (Figure~\ref{fig:ll2}). As the CDs profiles are different for different sources and different tracers, we model them using a Plummer-like function that describes flat plateaux and powerlaw decreasing at  higher column density as:
\begin{equation}
L(> L_{\rm bin}) = L_{\rm tot}\left( \frac{L_{\rm cut}}{\sqrt{L_{\rm 
cut}^2 + L_{\rm bin}^2 }} \right)^{p} ~~~.
\label{eq:plummer}
\end{equation}

$L_{\rm bin}$ is the luminosity threshold of each bin, $ L_{\rm tot}$ is the total luminosity of the cloud, $L_{\rm cut}$ is where the luminosity profile changes from power-law shape to flat plateau, and $p$  is the power-law index of the profile's section at the high luminosity tail. Figures~\ref{fig:ll2} show that each cloud can be characterized by parameters $L_{\rm tot}$, $L_{\rm cut}$, ${p}$ from the profile in Equation~\ref{eq:plummer}. This differentiation is potential applicable to differentiate different types of clouds and also its behaviour in different dense gas tracers. We will explore this topic further in the next paper.

\begin{discussion}

\discuss{Hacar} {You are comparing line luminosities but line tracers are subject
to excitation, opacity and temperature effects. How do you take into account all
these effects in your comparisons?}

\discuss{Quang} {The better way is to compare the column density distributions of
different lines if we can have multiple transitions of multiple isotopologue
observations of the same molecules. With only lines at single transitions, we can
only compare them and examine how excitation, opacity and temperature affect the
line luminosity distributions. }

\discuss{Marov} {In modern cosmology there are theories assuming dark matter
properties with an existence of multiple galaxies around the Milky Way core. Could
you associate dense clouds which you have been talking about with such faint
observing features?}

\discuss{Quang} {I couldn't comment on that because there is no observational
evidence to distinguish dark matter properties associated with multiple galaxies
in the Milky Way vicinity.}

\end{discussion}

\end{document}